\documentstyle[epsfigs,12pt,a4ps]{article}

\newcommand{\cb}{Crystal Barrel collaboration }

\newcommand{\etpi}{\mbox{$\eta\pi$ }}
\newcommand{\etpio}{\mbox{$\eta\pi^0$ }}
\newcommand{\etpim}{\mbox{$\eta\pi^-$ }}

\newcommand{\Mr}{\mbox{\rm M }}

\newcommand{\gams}{\mbox{$\gamma$s }}
\newcommand{\delpp}{\mbox{$\Delta^{++}$ }}
\begin{document}
\begin{titlepage}
\def\footnoterule{\hrule width 1.0\columnwidth}
\begin{tabbing}
put this on the right hand corner using tabbing so it looks
 and neat and in \= \kill
\> {10 July 2000}
\end{tabbing}
\bigskip
\bigskip
\begin{center}{\Large  {\bf A study of the centrally produced
\etpio and \etpim systems in $pp$ interactions at 450 GeV/c}
}\end{center}
\bigskip
\bigskip
\begin{center}{        The WA102 Collaboration
}\end{center}\bigskip
\begin{center}{
D.\thinspace Barberis$^{  4}$,
F.G.\thinspace Binon$^{   6}$,
F.E.\thinspace Close$^{  3,4}$,
K.M.\thinspace Danielsen$^{ 10}$,
S.V.\thinspace Donskov$^{  5}$,
B.C.\thinspace Earl$^{  3}$,
D.\thinspace Evans$^{  3}$,
B.R.\thinspace French$^{  4}$,
T.\thinspace Hino$^{ 11}$,
S.\thinspace Inaba$^{   8}$,
A.\thinspace Jacholkowski$^{   4}$,
T.\thinspace Jacobsen$^{  10}$,
G.V.\thinspace Khaustov$^{  5}$,
J.B.\thinspace Kinson$^{   3}$,
A.\thinspace Kirk$^{   3}$,
A.A.\thinspace Kondashov$^{  5}$,
V.N.\thinspace Kolosov$^{  5}$,
A.A.\thinspace Lednev$^{  5}$,
V.\thinspace Lenti$^{  4}$,
I.\thinspace Minashvili$^{   7}$,
J.P.\thinspace Peigneux$^{  1}$,
V.\thinspace Romanovsky$^{   7}$,
N.\thinspace Russakovich$^{   7}$,
A.\thinspace Semenov$^{   7}$,
P.M.\thinspace Shagin$^{  5}$,
H.\thinspace Shimizu$^{ 12}$,
A.V.\thinspace Singovsky$^{ 1,5}$,
A.\thinspace Sobol$^{   5}$,
M.\thinspace Stassinaki$^{   2}$,
J.P.\thinspace Stroot$^{  6}$,
K.\thinspace Takamatsu$^{ 9}$,
T.\thinspace Tsuru$^{   8}$,
O.\thinspace Villalobos Baillie$^{   3}$,
M.F.\thinspace Votruba$^{   3}$,
Y.\thinspace Yasu$^{   8}$.
}\end{center}

\begin{center}{\bf {{\bf Abstract}}}\end{center}

{
A partial wave analysis of the centrally produced \etpio
and \etpim channels has been performed in $pp$ collisions using
an incident beam momentum of 450~GeV/c.
Clear $a_0(980)$ and $a_2(1320)$ signals have been observed in $S$
and $D_+$ waves respectively.
The $dP_T$, $\phi$ and $|t|$ distributions of these resonances are
presented.
}
\bigskip
\bigskip
\bigskip
\begin{tabbing}
aba \=   \kill
$^1$ \> \small
LAPP-IN2P3, Annecy, France. \\
$^2$ \> \small
Athens University, Physics Department, Athens, Greece. \\
$^3$ \> \small
School of Physics and Astronomy, University of Birmingham, Birmingham, U.K. \\
$^4$ \> \small
CERN - European Organization for Nuclear Research, Geneva, Switzerland. \\
$^5$ \> \small
IHEP, Protvino, Russia. \\
$^6$ \> \small
IISN, Belgium. \\
$^7$ \> \small
JINR, Dubna, Russia. \\
$^8$ \> \small
High Energy Accelerator Research Organization (KEK), Tsukuba,
Ibaraki 305-0801, Japan. \\
$^{9}$ \> \small
Faculty of Engineering, Miyazaki University, Miyazaki 889-2192, Japan. \\
$^{10}$ \> \small
Oslo University, Oslo, Norway. \\
$^{11}$ \> \small
Faculty of Science, Tohoku University, Aoba-ku, Sendai 980-8577, Japan. \\
$^{12}$ \> \small
RCNP, Osaka University, Ibaraki, Osaka 567-0047, Japan. \\
\end{tabbing}
\end{titlepage}
\setcounter{page}{2}
\bigskip
\par
The $\eta\pi$ system has been studied in several experiments
in the search for exotic states~\cite{pwave,GAMS3}.
The interest in this system is mainly caused by the fact that
$P$-wave in the $\eta\pi$ system
carries exotic quantum numbers $J^{PC} = 1^{-+}$.
\par
In addition to the well-known $a_0(980)$ and $a_2(1320)$,
two new
$q\bar q$ states were observed recently
in the \etpi system.
In 1994 the \cb claimed a new $I^GJ^{PC}=1^-0^{++}$ state,
the $a_0(1450)$ \cite{CB3}, which they
then confirmed in different channels \cite{CB2,CB4,CB5}.
This
resonance could be considered as the isovector member of the
ground state scalar multiplet instead
of the $a_0(980)$, which is a peculiar object in meson spectroscopy and
whose nature is still not completely clear, see for example ref.~\cite{a0ref}.
In addition,
the \cb observed a new isovector tensor state, the $a_2(1660)$, decaying
to \etpio \cite{CB-a2}.
\par
In a previous analysis of the centrally produced \etpio system by the
NA12/2 collaboration~\cite{kinashi} clear $a_0(980)$ and $a_2(1320)$
signals were observed but
no evidence for a $J^{PC} = 1^{-+}$
state was found.
An interesting feature of that analysis
was the ratio of the $a_0(980)$ to $a_2(1320)$
which was found
to be much bigger than had been observed in $\pi^- p $
interactions~\cite{GAMS3}.
\par
This paper presents a study of the centrally produced $\eta\pi$ final state.
Firstly the reaction
\begin{equation}
pp \rightarrow p_{s} (\eta \pi^o) p_{f}
\label{eq:r1}
\end{equation}
has been studied at 450~GeV/c, where the $\eta$
is observed decaying to 2\gams.
The subscripts $f$ and $s$ indicate the
fastest and slowest particles in the laboratory respectively.
The WA102 experiment
has been performed using the CERN Omega Spectrometer,
the layout of which is described in ref.~\cite{WADPT}.
The GAMS-4000 multiphoton spectrometer has been used to detect \gams.
\par
Reaction (\ref{eq:r1}) has been isolated from the sample
of events having two outgoing charged tracks and four \gams
by first imposing the following cuts on the components of the
missing momentum:
$|$missing~$P_{x}| <  17.0$ GeV/c,
$|$missing~$P_{y}| <  0.16$ GeV/c and
$|$missing~$P_{z}| <  0.12$ GeV/c,
where the x axis is along the beam direction.
A correlation between
pulse-height and momentum
obtained from a system of
scintillation counters was used to ensure that the slow
particle was a proton.
\par
The separation of events belonging to reaction (\ref{eq:r1}) has been performed
on the basis of a kinematical analysis (6C fit, four-momentum
conservation being used and the masses of the two mesons being fixed).
There is evidence for a small
$\Delta^+(1232)$ signal in the $p_f\pi^0$ mass
spectrum which has been removed
by requiring $\Mr (p_f\pi^0)$~$>$~1.3~GeV.
Detector acceptances and efficiencies have been calculated using a Monte-Carlo
simulation taking into account the geometry of the
set up, detector resolutions,
event selection criteria and the kinematical fitting procedure.
The resulting centrally produced \etpio effective mass
distribution corrected for efficiency and rescaled to the
total number of observed events is shown
in fig.~\ref{fi:1}a) and consists of 6045 events.
There is clear
evidence for the $a_0(980)$ and $a_2(1320)$.
As can be seen from the $\gamma \gamma $ mass spectrum shown as an
inset (fig~\ref{fi:1}b)) there is approximately 30 \% background below
the $\eta$ signal.
To investigate the effects of this background
the side bands around the $\eta$ signal have been studied.
Superimposed on the mass spectrum as a shaded histogram is the
estimate of the background.
Fig.~\ref{fi:1}c) shows the background subtracted mass spectrum.

To determine the parameters of the
resonances  produced in  reaction (\ref{eq:r1}) a  fit to the
efficiency corrected mass spectrum has been performed using
a parametrisation of the form
$$
{dN}/{dm}\ (m)  =
           BG(m) +
           a_1 |BW_{a_0(980)}(m,0)|^2 +
           a_2 |BW_{a_2(1320)}(m,2)|^2
$$
where
$$
BG(m)  =  (m-m_{thr})^{\alpha}
e^{-\beta{m}-\gamma{m}^2}
$$
represents the background.
$m$ is the
\etpi
mass,
$m_{thr}$ is the
\etpi
threshold mass and
$a_n$,
$\alpha$, $\beta$ and $\gamma$ are parameters to be determined from the fit.
Relativistic Breit-Wigner functions $BW_i(m,J)$, where $J$ is the spin
of the resonance,
have been convoluted with a Gaussian to account for
the experimental mass resolution.
The Breit-Wigner used to describe the $a_0(980)$ uses the
Flatt\'{e}
formula~\cite{flatte}.
In order to describe the
centrally produced
$\eta \pi$ mass spectra
the differential cross section has been multiplied by the
kinematical factor $(M_{\eta\pi} - 4m_{thr}^2)^{1/2}/M_{\eta\pi}^3$~\cite{AMP}.
The fit is shown in fig.~\ref{fi:1}c) and
gives
$M(a_0(980))= 975 \pm 7$ MeV, $\Gamma(a_0(980)) = 72 \pm 16$ MeV and
$M(a_2(1320)=1308 \pm 9 $ MeV, $\Gamma(a_2(1320)) =115 \pm 20$  MeV.
The parameters found for the $a_0(980)$ and $a_2(1320)$ are consistent with
those from the PDG~\cite{PDG98}.
\par
A PWA has been performed in the mass interval from 670 to 2000 MeV for
the \etpio final state in 60 MeV mass bins.
The PWA of the centrally produced \etpi
system has been made assuming that the \etpi system is produced by the
collision of two particles (referred to as exchanged particles) emitted
by the scattered protons.
The z axis is defined by the momentum vector of the
exchanged particle with the greatest four-momentum transferred
in the \etpi centre of mass. The y axis is defined
by the cross product of the
momentum vectors of the two exchanged particles in the
$pp$ centre of mass. The two variables needed to specify the
decay process were taken as the polar
and azimuthal angles ($\theta$,$\phi$)=$\Omega$ of the $\pi$ in the \etpi
centre of mass relative to the coordinate system described above.
The amplitudes used for the PWA are defined in the reflectivity
basis~\cite{reflect}.
In this basis the angular distribution $I(\Omega)$ is given by a
sum of two non-interfering terms corresponding to negative and
positive values of reflectivity $\epsilon$. The waves used were
of the form $J_0$ ($m\epsilon=0-$), $J_-$ ($m\epsilon=1-$) and
$J_+$ ($m\epsilon=1+$). Only $S$, $P$ and $D$ waves
(corresponding to $J=0,1,2$)
have been used in the PWA with $m=0,1$. It has been determined from a
study of the moments that
the contributions of higher waves are negligible in the mass region under
study.
Finally, 12 parameters have to be determined from the fit
to the angular distributions.
In this model of the PWA the number of non-trivial solutions in
each mass interval can be less than or equal to 8 due to the ambiguities
problem \cite{reflect,sad8}.
In each mass bin an event-by-event maximum Likelihood
method has been used.
The PWA analysis has been performed by extending the
likelihood function to include the
background subtraction, namely the function
\[
F=-\sum_{i=1}^Nln\{I(\Omega)\} + \sum_{L,M}t_{LM}\epsilon_{LM}
+\sum_{i=1}^{N_{bg}}ln\{I(\Omega)\}
\]
has been minimised, where $N$ is the number of events in a given mass bin,
$\epsilon_{LM}$ are the efficiency corrections calculated
in the centre of the bin,
$t_{LM}$ are the moments of the angular distribution and
$N_{bg}$ is the number of background events.

All solutions for $S$, $P$ and $D$ waves have been found for
each interval.
In the mass region above 1.2 GeV, because the D-wave is the
dominant contribution, the solutions are not
affected by ambiguities and the 8 solutions are
effectively identical.
The threshold region does suffer from ambiguities, however,
we have picked the solution in which
the known $a_0(980)$ is in the S-wave.
\par
The physical solution for the \etpio system is
shown in fig.~\ref{fi:1}d). As can be seen
the $D_+$ wave dominates above 1.2 GeV.
The fit of the $D_+$ wave
amplitude squared with a Breit-Wigner function gives
parameters for $a_2(1320)$
meson similar to the ones from the fit to the efficiency
corrected mass spectrum.
A fit to the
$S$ wave amplitude squared gives
parameters for $a_0(980)$ similar to those from a
fit to the mass spectrum.
There is no evidence for an $a_0(1450)$ nor an $a_2(1660)$.
As can be seen
there is no evidence for other significant structures in any other wave.
\par
In previous analyses we have observed that I~=~1 mesons are suppressed
in the reaction $pp \rightarrow pp X^0$
but are enhanced in the reaction
$pp \rightarrow \Delta^{++} p X^-$~\cite{phiangpap}.
Therefore,
the reaction
\begin{equation}
pp \rightarrow p_{s} (\eta \pi^-) \Delta^{++}
\label{eq:r2}
\end{equation}
with the \delpp observed decaying to $p_{f} \pi^+$,
has been studied at 450~GeV/c. The $\eta$
is observed decaying to 2\gams.
Reaction (\ref{eq:r2}) has been isolated from the sample
of events having four outgoing charged tracks and two \gams
by imposing the cuts on the components of
missing momentum used for reaction (\ref{eq:r1}).
Events have been separated by a
kinematical analysis (5C fit). The events with a $\Delta^{++}(1232)$
in the final state have been selected by requiring
$\Mr (p_f\pi^+)$~$<$~1.4~GeV.
The background process
$pp \to p_{s} (\eta \pi^+ \pi^-) p_{f}$  gives the biggest
background below the \delpp peak.
By requiring $\Mr (\eta \pi^+ \pi^-)$~$>$~1.5~GeV
the signal to background in the \delpp region is greater than 10
(fig~\ref{fi:2}b)).
The resulting acceptance corrected \etpim mass spectrum
rescaled to the total number of observed events is shown
in fig.~\ref{fi:2}a) and consists of 8027 events.
The mass spectrum is similar to that observed in the \etpio final state
but the ratio $a_0$/$a_2$ is different.
In the $\gamma \gamma $ mass spectrum (not shown)
there is approximately 30 \% background below
the $\eta$ signal.
To investigate the effects of this background
the side bands around the $\eta$ signal have been studied.
Superimposed on the mass spectrum as a shaded histogram is the
estimate of the background.
\par
A fit to the centrally produced background subtracted
\etpim system is shown in
fig~\ref{fi:2}c).
The fit yields
$M(a_0(980))= 988 \pm 8$ MeV, $\Gamma(a_0(980)) = 61 \pm 19$  MeV and
$M(a_2(1320)=1316 \pm 9 $ MeV, $\Gamma(a_2(1320)) = 112 \pm 14$  MeV.
\par
A PWA has been performed in the mass interval from 670 to 2000 MeV for
the \etpim final state in 60 MeV mass bins using the method described above.
The physical solution for the \etpim system is
shown in fig.~\ref{fi:2}d). As can be seen
the $D_+$ wave dominates above 1.2 GeV, the fit of the $D_+$ wave
amplitude squared with a Breit-Wigner function gives the
parameters for $a_2(1320)$
meson similar to the ones from the fit to the efficiency
corrected mass spectrum.
A fit to the
$S$ wave amplitude squared gives the same
parameters for $a_0(980)$ similar to those from
the fit to the mass spectrum.
Again there is no evidence for an $a_0(1450)$ nor an $a_2(1660)$.
As can be seen
there is no evidence for other significant structures in any other wave.
\par
The fits to \etpio and \etpim mass spectra,
give the following values for the production ratio of the $a_0(980)$
and $a_2(1320)$ decaying to \etpi:
\[
\frac{\sigma (pp \rightarrow pp[a_0(980) \rightarrow \eta \pi])}
{\sigma (pp \rightarrow pp[a_2(1320) \rightarrow \eta \pi])}
= 2.0 \pm 0.3
\]
for reaction
(\ref{eq:r1}) and
\[
\frac{\sigma (pp \rightarrow p\Delta^{++}[a_0(980) \rightarrow \eta \pi])}
{\sigma (pp \rightarrow p\Delta^{++}[a_2(1320) \rightarrow \eta \pi])}
= 0.8 \pm 0.2
\]
for reaction
(\ref{eq:r2}).
In the charge exchange reaction ($\pi^-p \rightarrow \eta \pi^0n$),
where the $a_2(1320)$ meson dominates
the \etpio mass spectrum~\cite{GAMS3}:
\[
\frac{\sigma (\pi^-p \rightarrow [a_0(980) \rightarrow \eta \pi]n)}
{\sigma (\pi^-p \rightarrow [a_2(1320) \rightarrow \eta \pi]n)}
\approx 0.15
\]
at 38 GeV/c beam momentum.

\par
In previous WA102 analyses it has been observed that
centrally produced states have different $dP_T$ dependencies, where
$dP_T$ is the difference in the transverse momentum vectors of the
two exchange particles \cite{dpt}. The ratio of the production
cross section for $dP_T < 0.2$ GeV to $dP_T > 0.5$ GeV
is significantly different for $q \bar q$ states
and for the glueball candidates.
It has been observed that
all undisputed $q\bar q$ states which can be produced by DPE
have very small
values for this ratio ($\leq$ 0.1). States which cannot
be produced by DPE (with negative $G$ parity, for example)
have slightly higher values ($\approx$ 0.25). And all
non-$q\bar q$ candidates
$f_0(980)$, $f_0(1500)$, $f_0(1710)$ and $f_2(1950)$
have values of this ratio about 1.
A study of the
\etpio and \etpim systems, which can not be produced by DPE,
has been made as a function of $dP_T$.
The results are given in table~\ref{ta:dpt}.

In addition, an interesting effect has been observed in
the azimuthal angle $\phi$ which is defined as the angle between the $p_T$
vectors of the two outgoing protons~\cite{phiangpap}.
In order to determine the azimuthal angle $\phi$ for the $a_0(980)$ and
$a_2(1320)$
the \etpio and \etpim mass spectra have been fitted in 30
degree bins of $\phi$ with the parameters of the resonances fixed
to those obtained from the fit to the total data. The resulting
distributions for the $a_0^0(980)$ and
$a_2^0(1320)$
are shown in
fig.~\ref{fi:3} a) and b) respectively.
The distributions for the
$a_0^-(980)$ and $a_2^-(1320)$
are shown in
fig.~\ref{fi:4} a) and b) respectively.

In order to determine the
four momentum transfer dependence ($t$) of the
resonances observed,
the \etpio mass spectrum has been fitted
in 0.1 GeV$^2$ bins of $t$
with the parameters of the resonances fixed to those obtained from the
fits to the total data.
Fig.~\ref{fi:3}c) and d) shows the four momentum transfer from
one of the proton vertices for the $a_0^0(980)$ and $a_2^0(1320)$
respectively.
The distributions
have been fitted with a single exponential
of the form $exp(-b |t|)$ and the
values of $b$ found are
$b=6.2\pm 0.8$ for the $a_0^0(980)$ and
$b=8.8\pm 0.4$ for
the $a_2^0(1320)$.
\par
In order to determine the
four momentum transfer ($|t|$) at the beam-$\Delta^{++}$
vertex for the $a_0^-(980)$ and $a_2^-(1320)$
the \etpim mass spectrum has been fitted as described above.
The resulting distributions
are shown in fig.~\ref{fi:4}c) and d) for the
$a_0^-(980)$ and $a_2^-(1320)$ respectively.
The distributions
have been fitted with a single exponential
of the form $exp(-b |t|)$ and the
values of $b$ found are
$b=4.0\pm 0.3$ for the $a_0^-(980)$ and
$b=5.9\pm 0.5$ for
the $a_2^-(1320)$.
\par
The four momentum transfer ($|t_{slow}|$) distributions
at the target-slow
vertex
are shown in fig.~\ref{fi:4}e) and f) for the
$a_0^-(980)$ and $a_2^-(1320)$ respectively.
The data for $|t|$~$\leq$~0.1~GeV$^2$
has been excluded from the fit
due to the poor acceptance for the slow proton in this range.
The distributions
have been fitted with a single exponential
of the form $exp(-b |t|)$ and the
values of $b$ found are
$b=7.5\pm 0.4$ for the $a_0^-(980)$ and
$b=7.9\pm 0.6$ for
the $a_2^-(1320)$.
\par
In summary,
a partial wave analysis of the centrally produced \etpio
and \etpim channels has been performed in the mass region
670 $-$ 2000 MeV.
Clear $a_0(980)$ and $a_2(1320)$ signals are seen in the $S$ and $D_+$
waves respectively.
No evidence is found for any significant structures in other waves.
In particular, there is no evidence for the $a_0(1450)$ nor the $a_2(1660)$.
The ratio of the $a_0(980)$ to $a_2(1320)$
is found to depend on the production mechanism.

\newpage
\begin{center}
{\bf Acknowledgements}
\end{center}
\par
This work is supported, in part, by grants from
the British Particle Physics and Astronomy Research Council,
the British Royal Society,
the Ministry of Education, Science, Sports and Culture of Japan
(grants no. 1004100 and 07044098), the Programme International
de Cooperation Scientifique (grant no. 576)
and
the Russian Foundation for Basic Research
(grants 96-15-96633 and 98-02-22032).
\bigskip

\newpage
\def\auth#1 {{#1}, \ }
\def\authet#1 {{#1} {\it et al.,}\ }
\def\jetp#1#2#3 {{\rm JETP Lett.} {\bf #1} (#2) #3}
\def\nc#1#2#3   {{\rm Nuovo Cim.} {\bf #1} (#2) #3}
\def\lnc#1#2#3  {{\rm Lett. Nuovo Cim.} {\bf #1} (#2) #3}
\def\nim#1#2#3  {{\rm Nucl. Instr. Meth.} {\bf #1} (#2) #3}
\def\np#1#2#3   {{\rm Nucl. Phys.} {\bf #1} (#2) #3}
\def\pl#1#2#3   {{\rm Phys. Lett.} {\bf #1} (#2) #3}
\def\eul#1#2#3  {{\rm Europhys.Lett.} {\bf #1} (#2) #3}
\def\epj#1#2#3  {{\rm Eur. Phys.J.} {\bf #1} (#2) #3}
\def\prep#1#2#3 {{\rm Phys. Rep.} {\bf #1} (#2) #3}
\def\prev#1#2#3 {{\rm Phys. Rev.} {\bf #1} (#2) #3}
\def\prl#1#2#3  {{\rm Phys. Rev. Lett.} {\bf #1} (#2) #3}
\def\rmp#1#2#3  {{\rm Rev. Mod. Phys.} {\bf #1} (#2) #3}
\def\rpp#1#2#3  {{\rm Rep. Prog. Phys.} {\bf #1} (#2) #3}
\def\sjnp#1#2#3 {{\rm Sov. J. Nucl. Phys.} {\bf #1} (#2) #3}
\def\pan#1#2#3  {{\rm Phys. Atom. Nucl. } {\bf #1} (#2) #3}
\def\yad#1#2#3  {{\rm Yad. Phys.} {\bf #1} (#2) #3}
\def\fiel#1#2#3 {{\rm Fiz.Elem.Chast.Atom.Yadra} {\bf #1} (#2) #3}
\def\spj#1#2#3  {{\rm Sov. Phys. JEPT} {\bf #1} (#2) #3}
\def\spu#1#2#3  {{\rm Sov. Phys.-Usp.} {\bf #1} (#2) #3}
\def\ufn#1#2#3  {{\rm Usp.Fys.Nauk} {\bf #1} (#2) #3}
\def\zp#1#2#3   {{\rm Zeit. Phys.} {\bf #1} (#2) #3}
\def\conf#1#2#3 {{\rm #1}, (#2) #3}
%


\newpage
\begin{table}[h]
\caption{Production of the resonances as a function of $dP_T$
expressed as a percentage of their total contribution and the
ratio (R) of events produced at $dP_T$~$\leq$~0.2~GeV to the events
produced at $dP_T$~$\geq$~0.5~GeV.}
\vspace{.3cm}
\label{ta:dpt}
\begin{center}
\begin{tabular}{|c|c|c|c|c|}
\hline
& & & & \\
Resonance &$dP_T$$\leq$0.2 GeV & 0.2$\leq$$dP_T$$\leq$0.5 GeV
&$dP_T$$\geq$0.5 GeV & $R=\frac{dP_T \leq 0.2 GeV}{dP_T\geq 0.5 GeV}$\\
& & & & \\
\hline
& & & & \\
$a^0_0(980)$  & 25$\pm$3 &33$\pm$5 &42$\pm$4 &  $0.57\pm0.09$ \\
& & & & \\
\hline
& & & & \\
$a^-_0(980)$  & 14$\pm$3 &37$\pm$2 &49$\pm$2 &  $0.29\pm0.06$ \\
& & & & \\
\hline
& & & & \\
$a^0_2(1320)$  & 10$\pm$2 &38$\pm$2 &52$\pm$3 &  $0.19\pm0.04$ \\
& & & & \\
\hline
& & & & \\
$a^-_2(1320)$  &  9$\pm$3 &39$\pm$2 &52$\pm$2 &  $0.17\pm0.06$ \\
& & & & \\
\hline
\end{tabular}
\end{center}
\end{table}

\clearpage
{ \large \bf Figures \rm}
\begin{figure}[h]
\caption{
a) The efficiency corrected mass spectrum
for the centrally produced \etpio events.
Superimposed as a shaded histogram is an estimation of the background
contribution.
Inset b) the $\gamma \gamma $ mass spectrum.
c) The efficiency corrected background subtracted mass spectrum
for the centrally produced \etpio events.
The curve is the result of the fit
described in the text.
d) The physical solutions from the PWA of
the \etpio events.
}
\label{fi:1}
\end{figure}
\begin{figure}[h]
\caption{
a) The efficiency corrected mass spectra
of centrally produced \etpim events.
Inset b) presents the
distribution to the $p_{f} \pi^+$ invariant masses.
c) The efficiency corrected background subtracted mass spectrum
for the centrally produced \etpim events.
The curve is the result of the fit
described in the text.
d) The physical solutions from the PWA of
the \etpim events.
}
\label{fi:2}
\end{figure}
\begin{figure}[h]
\caption{
The azimuthal angle $\phi$ between the $p_T$
vectors of the slow and fast protons for
a) the $a_0^0(980)$  and b) the $a_2^0(1320)$.
The four momentum transfer distributions for
c) the $a_0^0(980)$  and d) the $a_2^0(1320)$,
with the fits to the form $e^{-b|t|}$.
}
\label{fi:3}
\end{figure}
\begin{figure}[h]
\caption{
The azimuthal angle $\phi$ between the $p_T$
vectors of the slow proton and fast $\Delta^{++}$ for
a) the $a_0^-(980)$ and b) the $a_2^-(1320)$.
The four momentum transfer distributions from
the beam-fast vertex for c)
the $a_0^-(980)$ and
d) the $a_2^-(1320)$ and for the target-slow vertex
e) the $a_0^-(980)$ and
f) the $a_2^-(1320)$,
with the fits to the form $e^{-b|t|}$.
}
\label{fi:4}
\end{figure}
\addtocounter{figure}{-4}
\newpage
\begin{figure}[h]
\begin{center}
\epsfig{figure=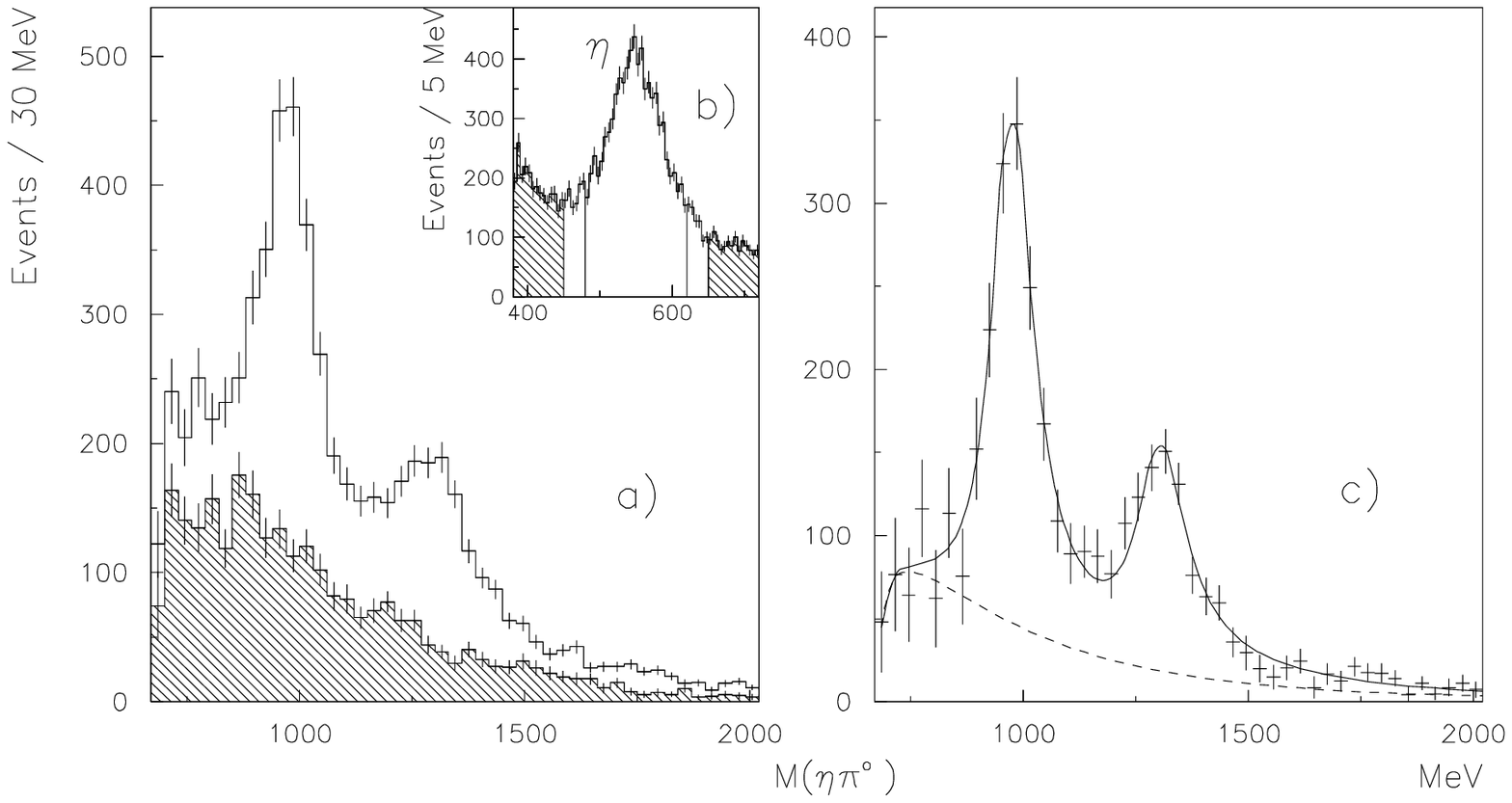,width=17cm,bbllx=0pt,bblly=0pt,bburx=567pt,bbury=367pt}
\epsfig{figure=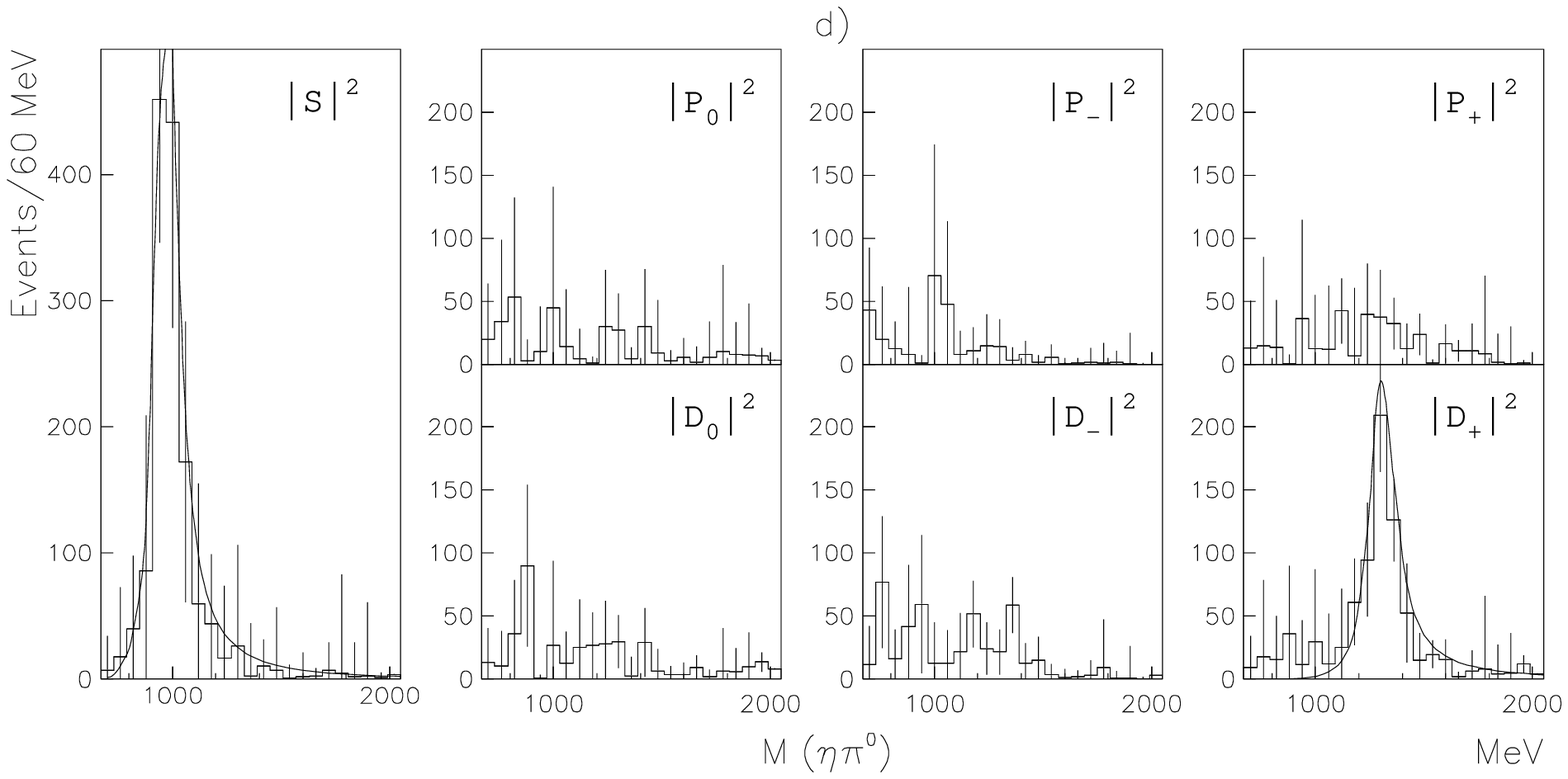,width=17cm,bbllx=0pt,bblly=-267pt,bburx=567pt,bbury=300pt}
\vspace{-70mm}
\caption{    }
\end{center}
\end{figure}
\newpage
\begin{figure}[h]
\begin{center}
\epsfig{figure=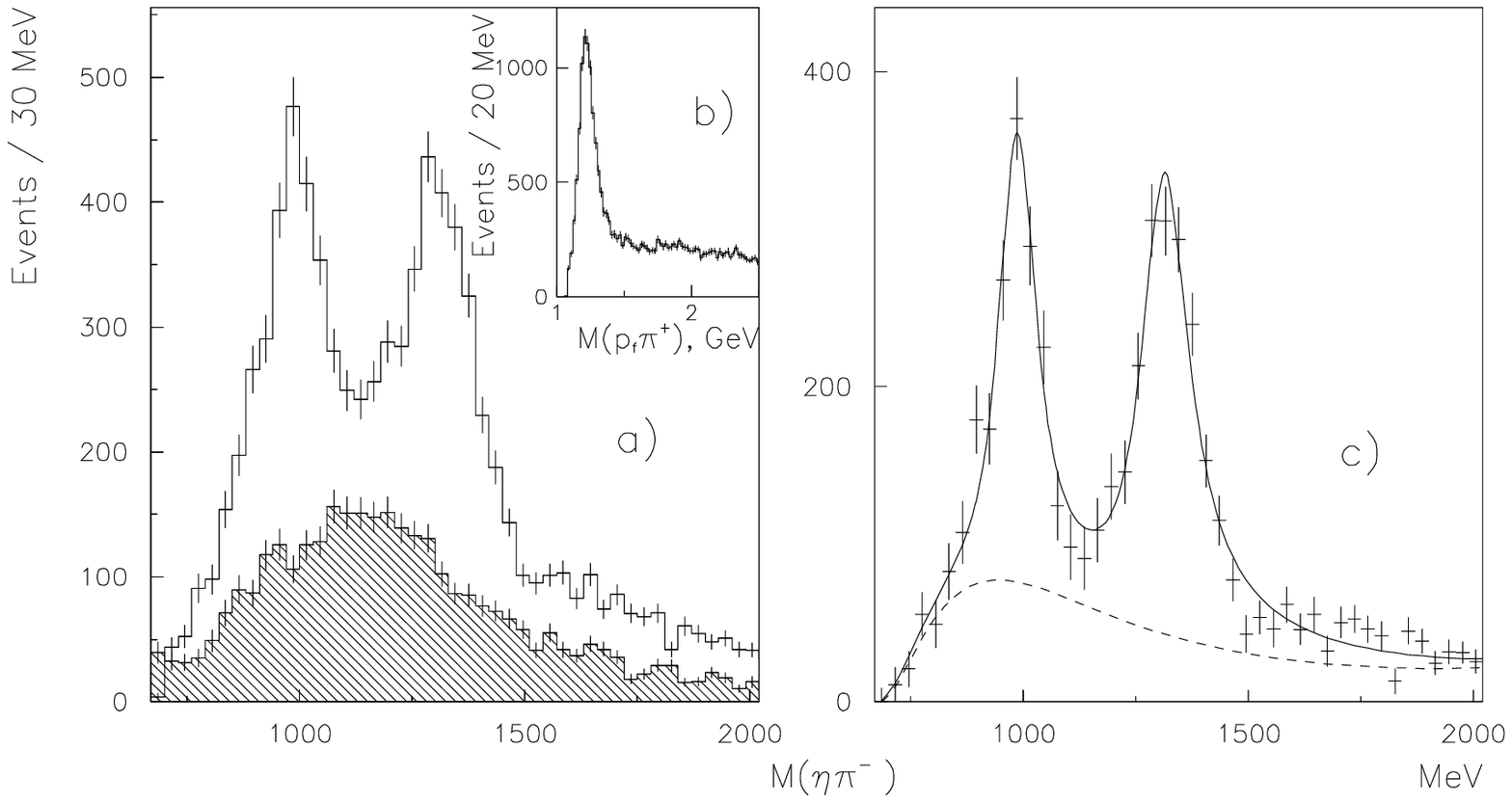,width=17cm,bbllx=0pt,bblly=0pt,bburx=567pt,bbury=367pt}
\epsfig{figure=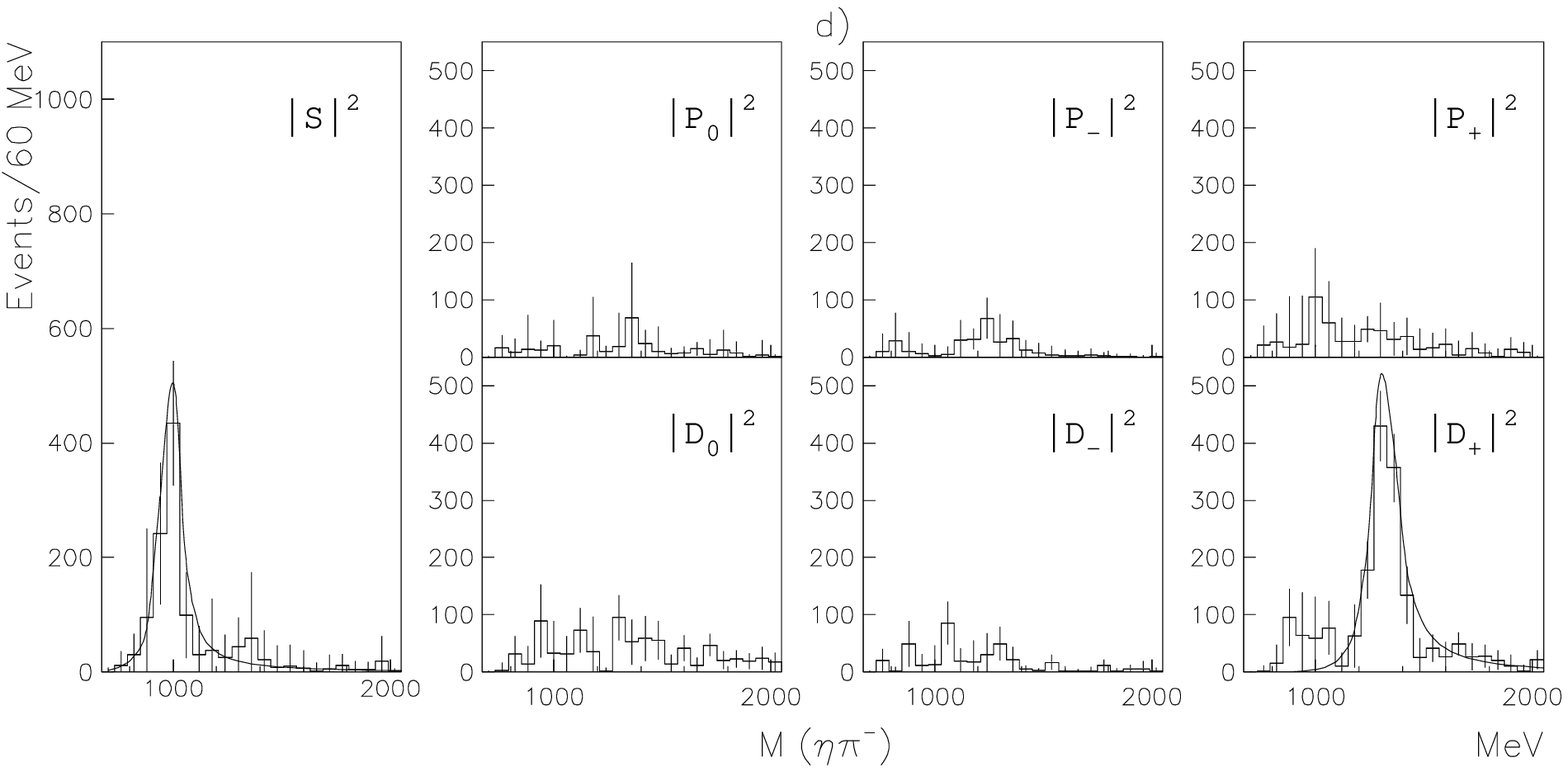,width=17cm,bbllx=0pt,bblly=-267pt,bburx=567pt,bbury=300pt}
\end{center}
\vspace{-70mm}
\caption{    }
\end{figure}
\newpage
\begin{figure}[h]
\begin{center}
\epsfig{figure=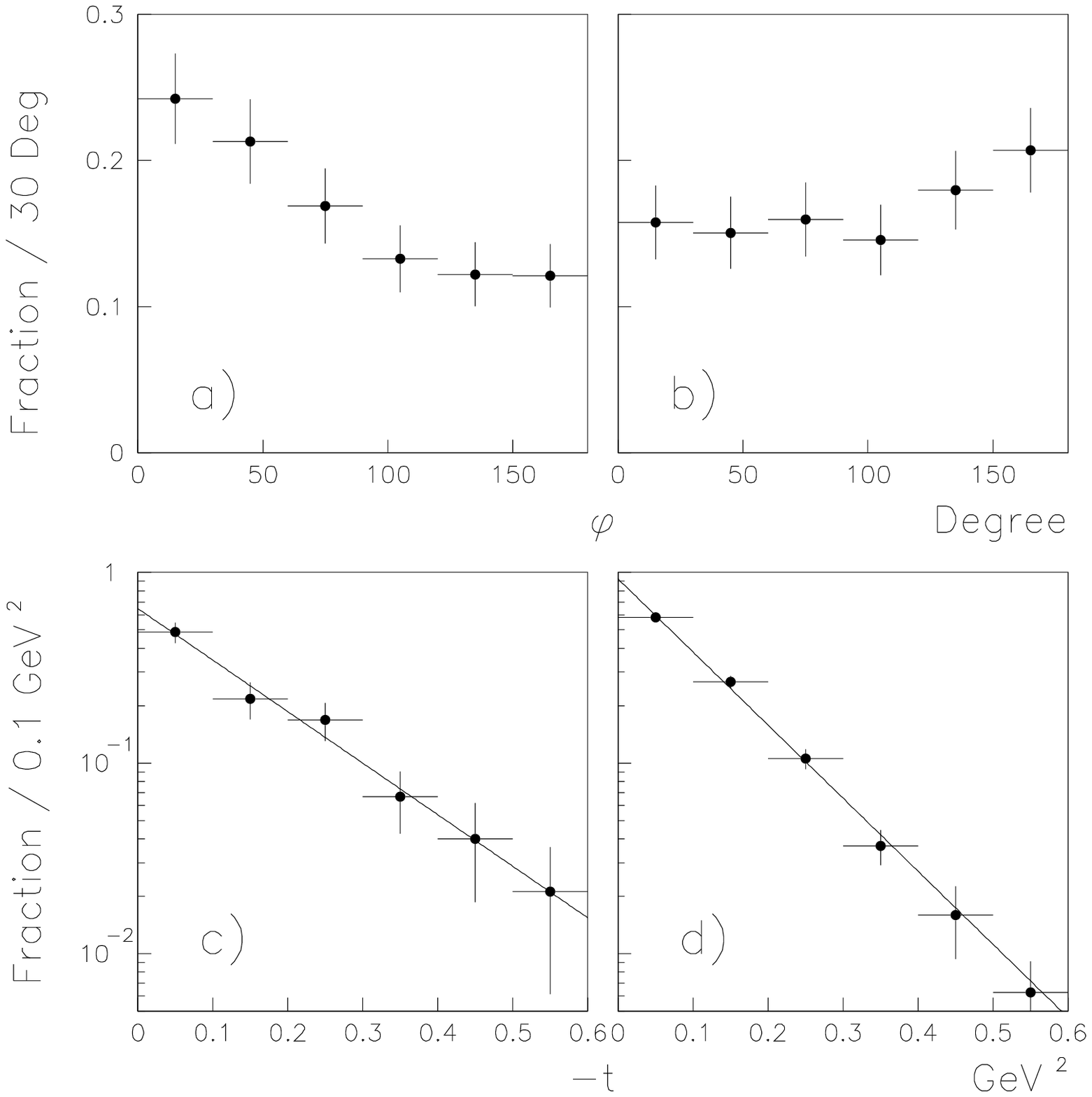,height=20cm,bbllx=40pt,bblly=0pt,bburx=567pt,bbury=567pt}
\end{center}
\vspace{-8mm}
\caption{    }
\end{figure}
\newpage
\begin{figure}[h]
\begin{center}
\epsfig{figure=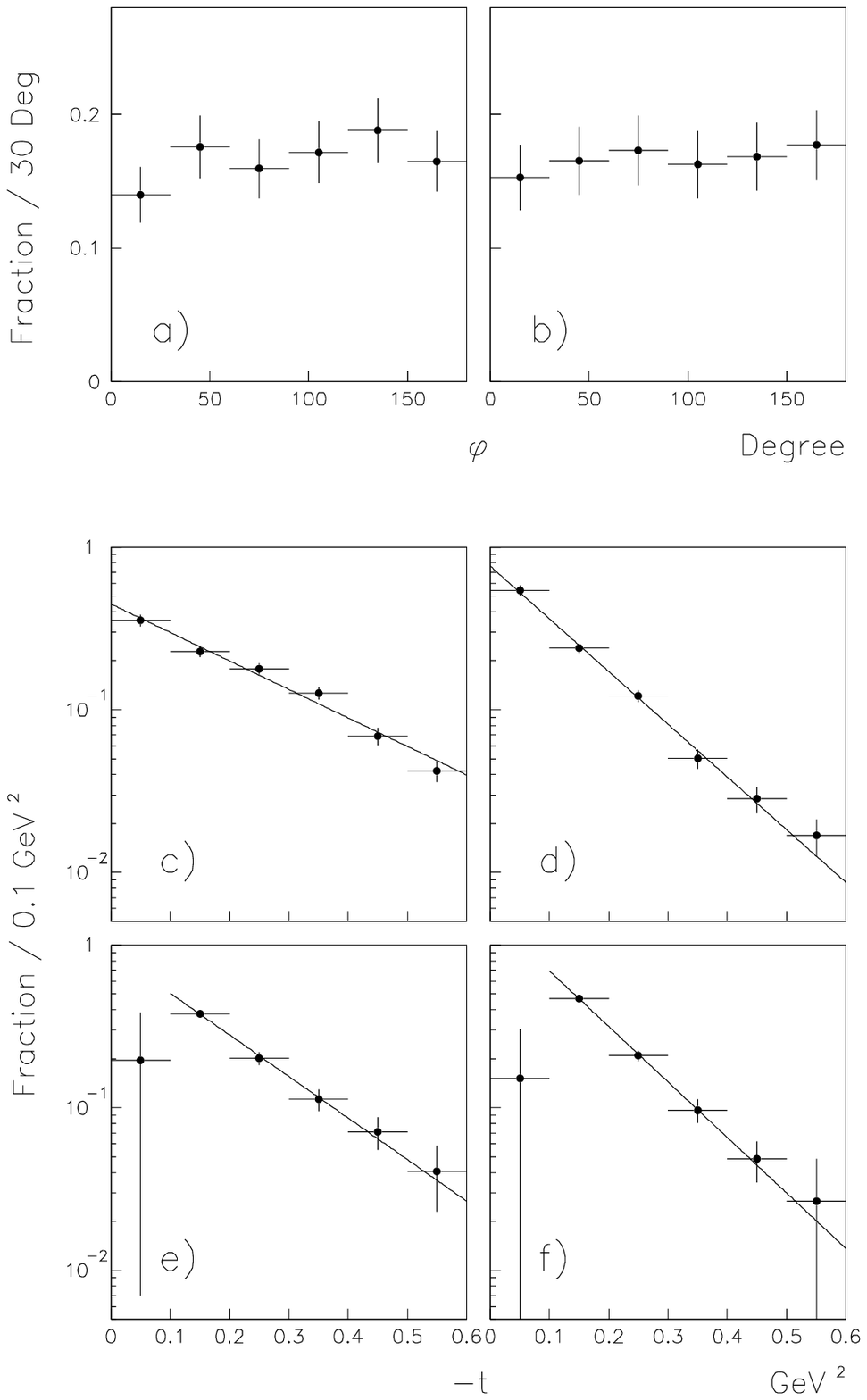,height=20cm,bbllx=-50pt,bblly=0pt,bburx=567pt,bbury=567pt}
\end{center}
\vspace{-8mm}
\caption{    }
\end{figure}

\begin{thebibliography}{99}

\bibitem{pwave}
\authet{ D.Alde}
\pl{B205}{1988}{397}; \\
\authet{ H.Aoyagi}
\pl{B205}{1988}{397};\\
\authet{ G.M.Beladidze}
\pl{B313}{1993}{276};\\
\authet{ D.R. Thompson}
\prl{79}{1997}{1630};\\
\authet{ A.Abele}
\pl{B423}{1998}{175}; \\
\authet{A.Abele}
\pl{B446}{1999}{275}.

\bibitem{GAMS3}
\authet{ D.Alde}
\pan{62}{1999}{421},
{}~\yad{62}{1999}{462}.

\bibitem{CB3}
\authet{C. Amsler}
\pl{B333}{1994}{277}.
\bibitem{CB2}
\authet{C. Amsler}
\pl{B446}{1999}{275}.


\bibitem{CB4}
\authet{C. Amsler}
\pl{B355}{1995}{425}.

\bibitem{CB5}
\authet{A.Abele}
\pl{B404}{1997}{179}.

\bibitem{a0ref}
\auth{R.L.Jaffe}
\prev{D15}{1977}{267}. \\
\auth{ L.Montanet}
\rpp{46}{1983}{337}. \\
\authet{N.N.Achasov}
\ufn{142}{1984}{361}. \\
\auth{F.E.Close}
\rpp{51}{1988}{833}. \\
\auth{J.Weinstein and N.Isgur}
\prev{D41}{1990}{2236}. \\
\authet{S.Ishida}
KEK Preprint 95-167,
\conf{Talk at HADRON'95, Manchester}{1995}{454}.

\bibitem{CB-a2}
\authet{ A.Abele}
\epj{C8}{1999}{67}.


\bibitem{kinashi}
\authet{ D.Alde}
KEK Preprint 95-160,
\conf{Talk at HADRON'95, Manchester}{1995}{448}.

\bibitem{WADPT}
\authet{D. Barberis}
\pl{B397}{1997}{339}.

\bibitem{flatte}
\authet{S. Flatt\'e}
\pl{B63}{1976}{224}.

\bibitem{AMP}
\auth{K.L. Au, D. Morgan and M.R. Pennington} \prev{D35} {1987}{1633}.
\bibitem{PDG98}
\auth{Particle Data Group} \epj{C3} {1998}{1}.

\bibitem{reflect}
\auth{S.U.Chung}
\prev{D56}{1997}{7299}.

\bibitem{sad8}
\auth{S.A.Sadovsky}
Preprint IHEP 91-71, Protvino, 1991.

\bibitem{phiangpap}
\authet{D.Barberis}
\pl{B467}{1999}{165}.
\bibitem{dpt}
\authet{D.Barberis}
\pl{B397}{1997}{339}.







\end{thebibliography}
\end{document}